\documentclass[twocolumn]{revtex4}

\usepackage{bm}

\usepackage{graphicx}

\usepackage{subfigure}
\newcommand{\ket}[1]{\left| #1 \right\rangle}

\newcommand{\cre}[2]{\hat{#1}^{\dag}_{#2}}
\newcommand{\ann}[2]{\hat{#1}_{#2}}

\begin{document}

\title{Einstein-Podolsky-Rosen Correlations of Ultracold Atomic Gases}

\author{Nir Bar-Gill$^1$}
\author{Christian Gross$^2$}
\author{Gershon Kurizki$^1$}
\author{Igor Mazets$^3$}
\author{Markus Oberthaler$^2$}

\affiliation{$^1$Weizmann Institute of Science, Rehovot, Israel.}
\affiliation{$^2$Kirchhoff-Institut f\"{u}r Physik, Universit\"{a}t Heidelberg, 69120 Heidelberg, Germany.}
\affiliation{$^3$Atominstitut \"{o}sterreichischer Universit\"{a}ten, TU Wien, Vienna, Austria.}

\begin{abstract}
Einstein, Podolsky \& Rosen (EPR) pointed out \cite{epr} that correlations induced between quantum objects will persist after these objects have ceased to interact. Consequently, their joint continuous variables (CV), e.g., the difference of their positions and the sum of their momenta, may be specified, regardless of their distance, with arbitrary precision. EPR correlations give rise to two fundamental notions\cite{mann,bell,schrodinger,peres,braunstein_rev}: {\em nonlocal ``steering''} of the quantum state of one object by measuring the other, and inseparability ({\em entanglement}) of their quantum states. EPR entanglement is a resource of quantum information (QI)\cite{braunstein_rev,drummond_reid,zoller_cve} and CV teleportation of light\cite{vaidman,braunstein} and matter waves\cite{opatrny,opatrny_deb}. It has lately been demonstrated for {\em collective} CV of distant thermal-gas clouds, correlated by interaction with a common field \cite{polzik1999,polzik_nature}. Here we demonstrate that collective CV of two species of trapped ultracold bosonic gases can be EPR-correlated (entangled) via {\em inherent} interactions between the species. This paves the way to further QI applications of such systems, which are atomic analogs of coupled superconducting Josephson Junctions (JJ)\cite{smerzi_prl,oberthaler}. A precursor of this study has been the observation of quantum correlations (squeezing) in a single bosonic JJ \cite{oberthaler_squeezing}.
\end{abstract}

\maketitle

%\spacing{1.5}

\paragraph*{EPR criteria --}
In studying continuous variable entanglement (CVE), it is instructive to draw an analogy with the original EPR scenario \cite{epr}, wherein two particles, $1$ and $2$, are defined through their position and momentum variables $x_{1,2},p_{1,2}$. EPR saw as paradox the fact that depending on whether we measure $x_1$ or $p_1$ of particle $1$, one can predict the measurement result of $x_2$ or $p_2$, respectively, with arbitrary precision, unlimited by the Heisenberg relation $\Delta x_2 \Delta p_2 \leq 1/2$ (choosing $\hbar=1$). 
This {\em nonlocal} dependence of the measurement results of particle 2 on those of particle 1 has been dubbed {\em ``steering''} by Schr\"odinger\cite{schrodinger2}.
Equivalently, the EPR state is deemed entangled in the continuous translational variables of the two particles.
The entanglement is exhibited by the collective operators $\hat{x}_{\pm} = \hat{x}_1 \pm \hat{x}_2$ and $\hat{p}_{\pm} = \hat{p}_1 \pm \hat{p}_2$. 
In quantum optics these variables are associated with the sum and difference of field quadratures of two light modes mixed by a symmetric beam splitter \cite{braunstein_rev,reid_rev} (Fig. 1a).

In order to quantify the EPR correlations, one may adopt two distinct criteria. The first criterion imposes an upper bound on the product of the variances of EPR-correlated {\em commuting} dimensionless operators, $\hat{x}_+$ and $\hat{p}_-$ or $\hat{x}_-$ and $\hat{p}_+$ \cite{drummond_reid,opatrny}:
\begin{equation}
\langle \Delta \hat{x}_{\pm}^2 \rangle \langle \Delta \hat{p}_{\mp}^2 \rangle \equiv \frac{1}{4s} \leq \frac{1}{4}, \label{eq:2mode_crit}
\end{equation}
%where we have used dimensionless variables, normalized such that the single-mode Heisenberg (shot-noise) uncertainty limit is $\frac{1}{4}$. 
The EPR correlation is then measured by the two-mode squeezing factor $\infty > s > 1$.
The second is the inseparability (entanglement) criterion for {\em gaussian} states \cite{zoller_cve,polzik_nature}, related to the sum of the variances of the correlated observables $\epsilon \equiv \langle \Delta \hat{x}_{\pm}^2 \rangle + \langle \Delta \hat{p}_{\mp}^2 \rangle - 1 < 0$.
Here the maximal entanglement corresponds to the most negative $\epsilon$ obtainable.
In what follows we inquire: to what extent do these EPR criteria apply to the system at hand, i.e., a two-species BEC in a symmetric double-well potential?

\paragraph*{Scheme for global-mode EPR correlations in bosonic JJs --}
We first consider the correlation dynamics of the two species (two internal states of the atom), in the presence of tunnel coupling between the wells. We shall analyze the EPR correlations in the basis of two global internal-state modes that are {\em not} spatially separated between the two wells. 
%Each modes pertains to a different internal state, and the relevant quadratures are the number and phase differences between the left and right wells (for each internal state).

Since there is no population exchange between the internal states $\ket{A}$ and $\ket{B}$, the numbers of atoms $N_A$ and $N_B$ in these states are constants of motion. The Hamiltonian (Supplement) can be then written in this basis in terms of the left-right atom-number differences in the two internal states, $\hat{n}_A = \left( \cre{a}{L} \ann{a}{L} - \cre{a}{R} \ann{a}{R} \right)/2$ and $\hat{n}_B = \left( \cre{b}{L} \ann{b}{L} - \cre{b}{R} \ann{b}{R} \right)/2$, and their canonically conjugate phase operators $\hat{\phi}_{A,B}$, obeying the commutation relations $\left[ \hat{\phi}_{\alpha} , \hat{n}_{\alpha'} \right] = i \delta_{\alpha \alpha'}$ ($\alpha,\alpha'=A,B$).
%In terms of these operators the Hamiltonian (\ref{BH2S}) becomes:
%\begin{eqnarray}
%H &=& E_{c11} \hat{n}_1^2 + E_{c22} \hat{n}_2^2 + 2 E_{c12} \hat{n}_1 \hat{n}_2 \nonumber \\
%&-& 2J \sqrt{\frac{N_1^2}{4} - \hat{n}_1^2} cos \hat{\phi}_1 - 2J \sqrt{\frac{N_2^2}{4} - \hat{n}_2^2} cos \hat{\phi}_2. \label{Hnphi}
%\end{eqnarray}
For simplicity we assume from now on that $N_A=N_B \equiv N$ (generalization to $N_A \neq N_B$ is straightforward), and consider small interwell number differences such that $\langle \hat{n}_{A,B} \rangle << N$. The Hamiltonian\cite{leggett_jphysb} then becomes
\begin{eqnarray}
H &=& (E_c)_{AA} \hat{n}_A^2 + (E_c)_{BB} \hat{n}_B^2 + 2 (E_c)_{AB} \hat{n}_A \hat{n}_B \nonumber \\
&-& JN \left( cos \hat{\phi}_A + cos \hat{\phi}_B \right) + \frac{2J}{N} \left( \hat{n}_A^2 cos \hat{\phi}_A + \hat{n}_B^2 cos \hat{\phi}_B \right). \label{Hnphi2}
\end{eqnarray}
Here the nonlinearity coefficients (``charging'' energies) $(E_c)_{AA}$, $(E_c)_{BB}$ and $(E_c)_{AB}$ are determined respectively by
%$E_{c \alpha \alpha'} = 4 \pi m^{-1} a_{\alpha \alpha'} \int d^3 r |\Phi_0 (r) |^4$, $m$ being the atomic mass, $a_{\alpha \alpha'}$ ($\alpha=1,2$) and $a_{12}$ being 
the intra- and inter-species s-wave scattering lengths. 
%The single-well ground-state wave function $\Phi_0 (r)$ (normalized to 1) and 
The tunneling energy $J$ is the same for the atoms in the internal states $\ket{A}$ and $\ket{B}$.%, subject to an identical potential.

\begin{figure}
\subfigure{
\includegraphics[width=0.48\linewidth,height=0.25\linewidth]{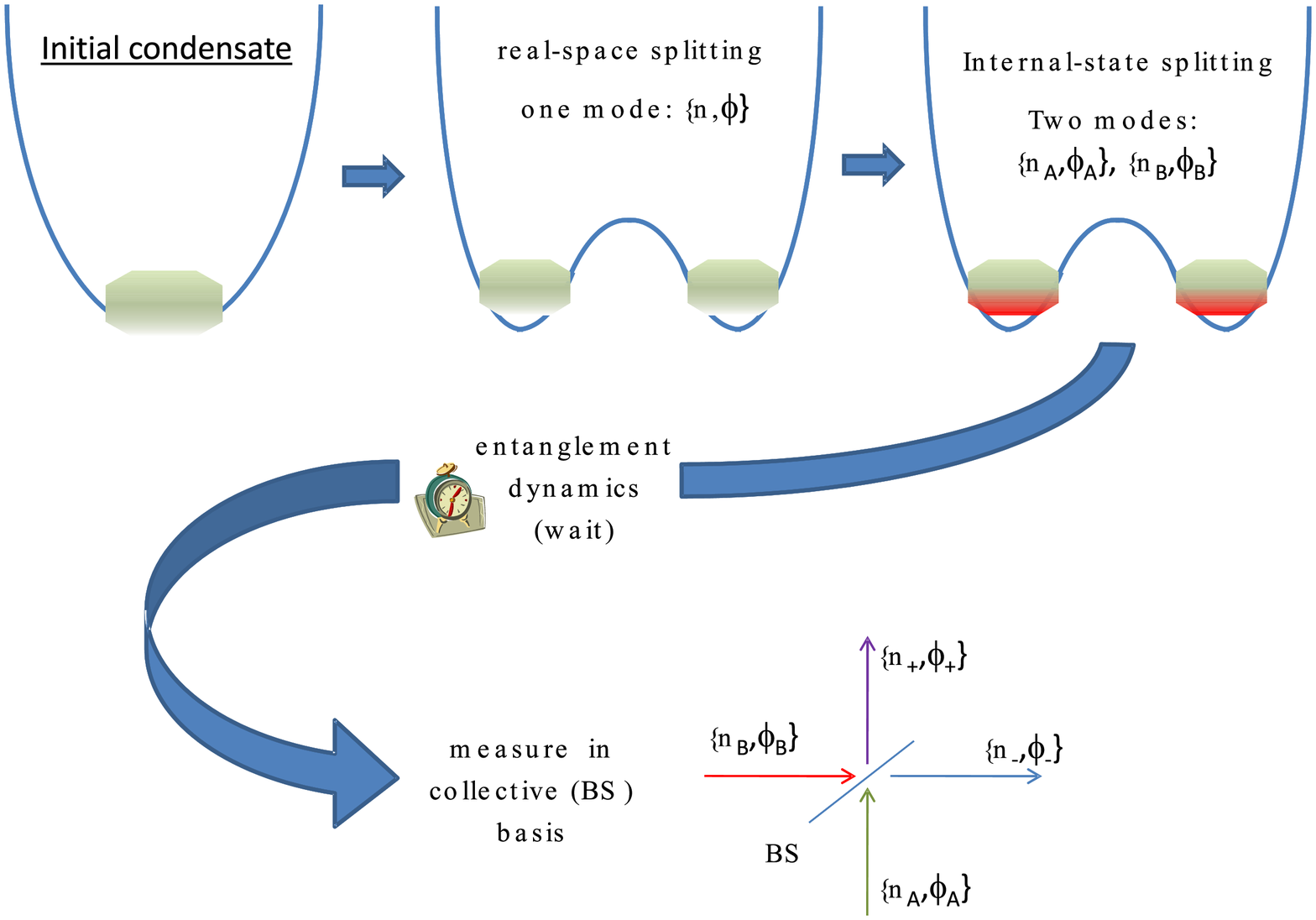}}
\subfigure{
\includegraphics[width=0.48\linewidth,height=0.25\linewidth]{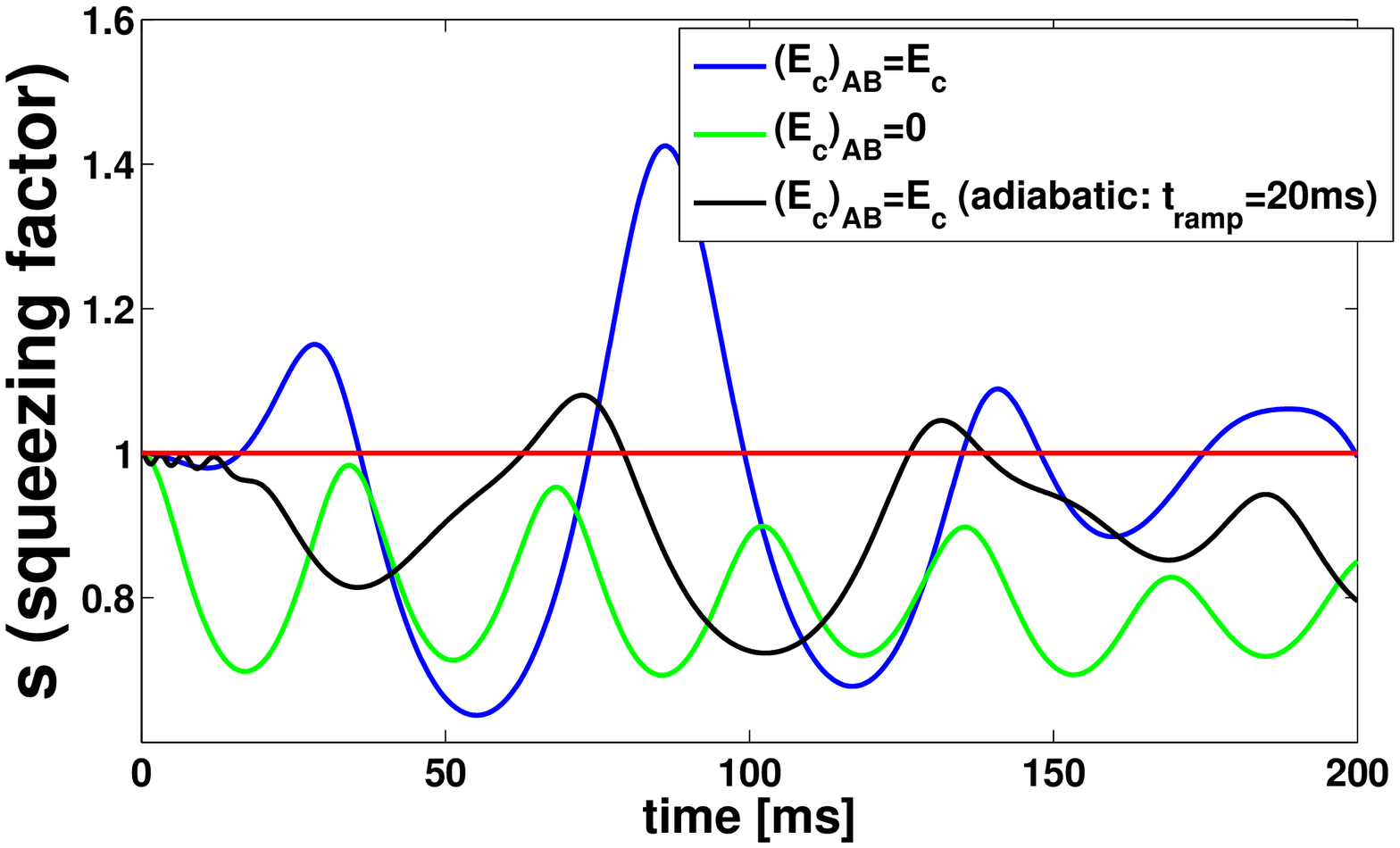}}
\subfigure{
\includegraphics[width=0.48\linewidth,height=0.25\linewidth]{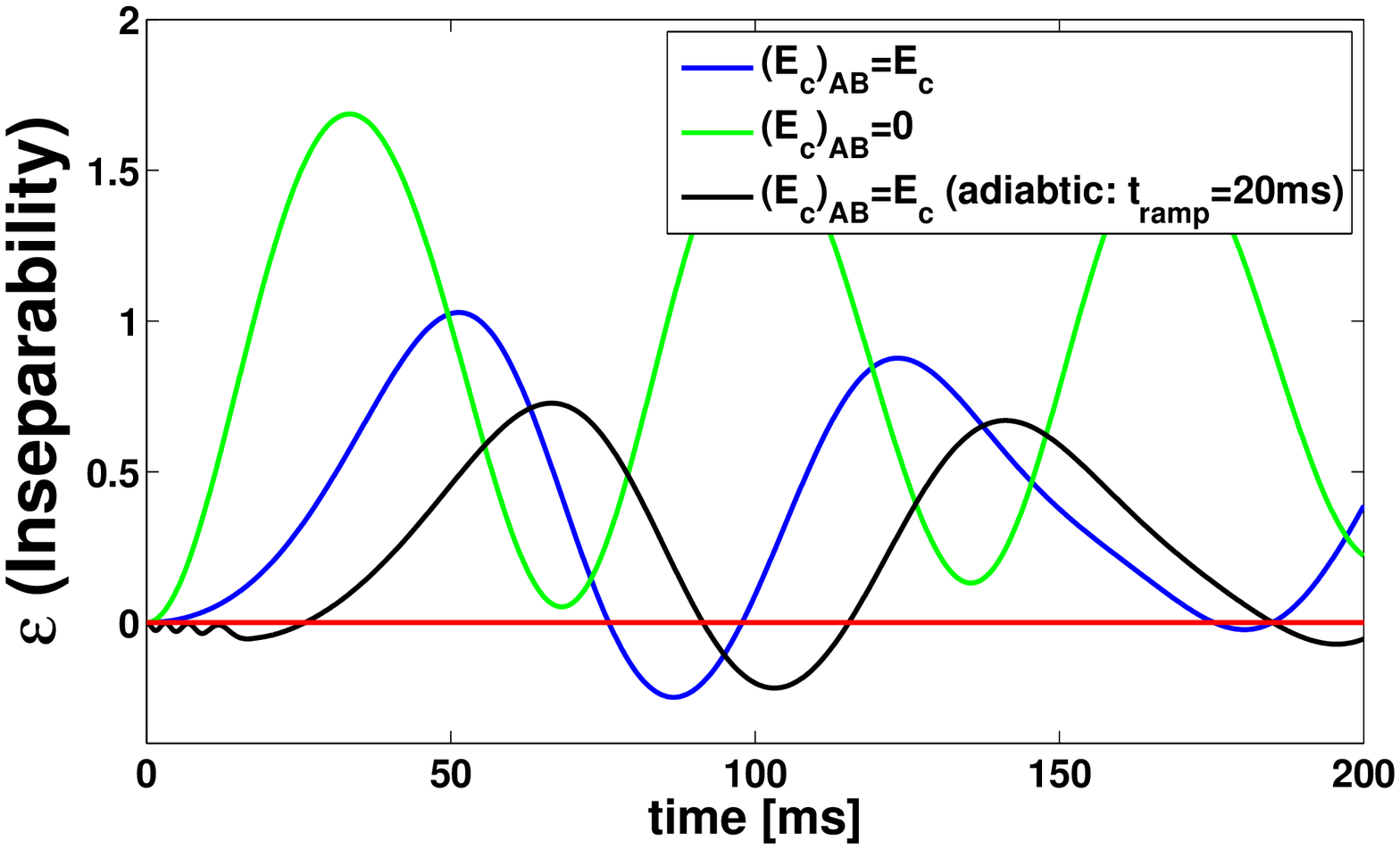}}
\subfigure{
\includegraphics[width=0.48\linewidth,height=0.25\linewidth]{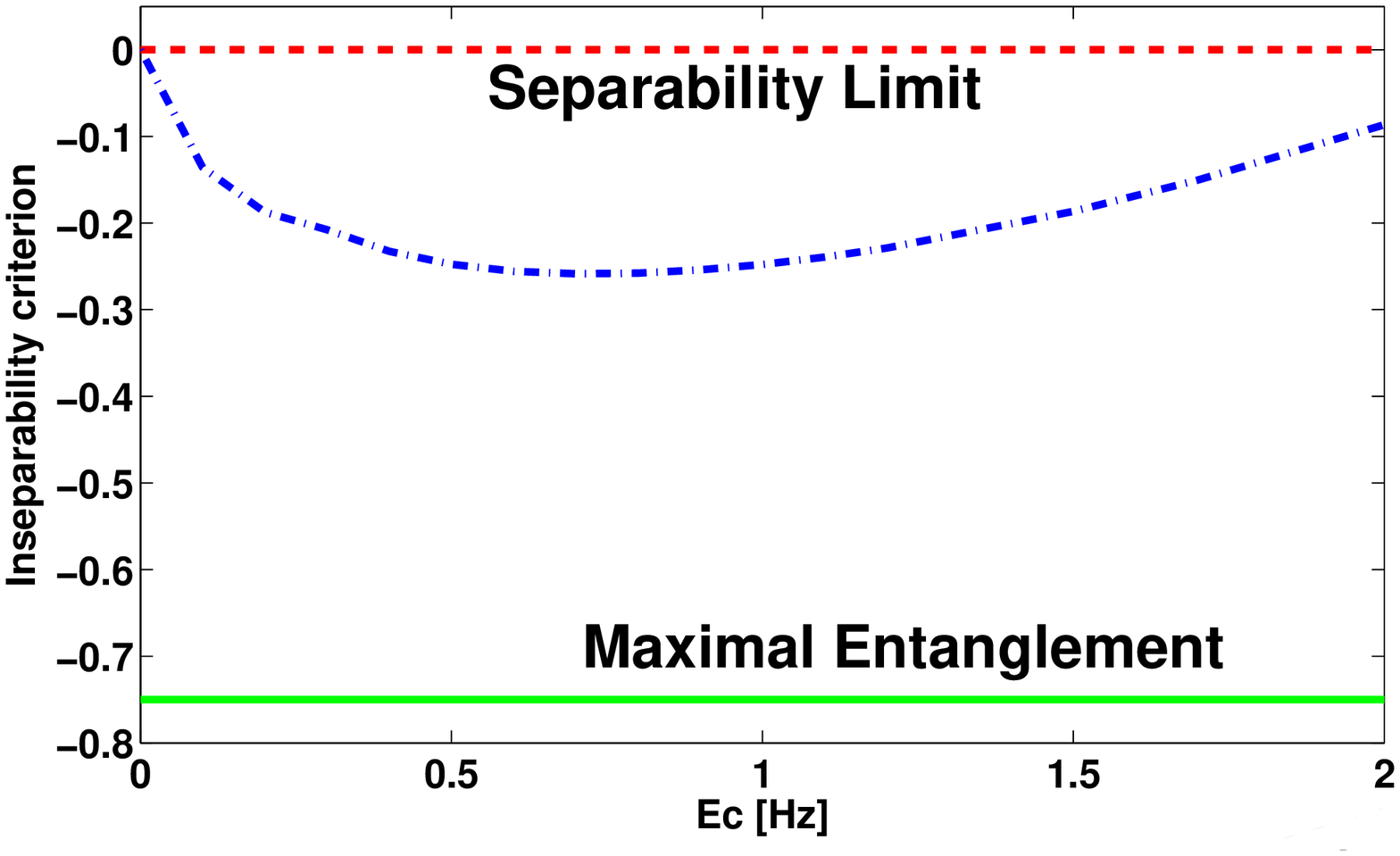}}
\caption{(a) State preparation scheme: the condensate is split both in real-space and in the internal-state basis, to create two-coupled modes. Then entanglement dynamics take place as a function of time, and the measurement is done in the collective beam-splitter basis. 
(b)-(c) Dynamics of the entanglement defined by EPR criteria for both {\em sudden} and slow intermode coupling barrier ramping-up, found through exact simulation of Eq. (1) in the Supplement. In each figure we plot the case of a coupled system ($E_{c12}=E_c$, solid blue), an uncoupled system ($E_{c12}=0$, solid green), slow ramp-up (solid black) and the classical limit (solid red).
The tradeoff between entanglement and nonlinear phase diffusion is better for the sudden coupling.
(d) Maximal amount of entanglement (blue dash-dotted line) reached through the dynamics, measured by the inseparability criterion in Eq. (\ref{L_crit}). We plot the maximal inseparability as a function of the charging energy, assuming that all coefficients are equal $(E_c)_{AA} = (E_c)_{BB} = (E_c)_{AB} \equiv E_c$\cite{greene,rbfeshbach}. It can be seen that for small charging energy $E_c$ the entanglement grows, but stronger interactions cause significant nonlinear phase diffusion, and therefore reduce the EPR correlations. From the competition between charging-induced entanglement and charging-induced nonlinear phase diffusion, we find the charging energy which gives maximal entanglement (see Methods). Red-dashed line indicates the separability limit, green-solid line indicates maximal inseparability for the ideal case (see text).
} \label{fig:squeezing}
\end{figure}

Equation (\ref{Hnphi2}) displays the full dynamics used in our numerics (Fig. \ref{fig:squeezing}), that of two quantum nonlinear pendula coupled via $2 (E_c)_{AB} \hat{n}_A \hat{n}_B$. This coupling is the key to EPR correlations of modes $A$ and $B$.

We may, for didactic purposes, simplify (\ref{Hnphi2}) by expanding the cosine terms. 
%The first term in the expansion gives
%\begin{eqnarray}
%H &=& \left( E_c + \frac{2J}{N} \right) \left( \hat{n}_1^2 + \hat{n}_2^2 \right) + 2 E_{c12} \hat{n}_1 \hat{n}_2 \nonumber \\
%&+& \frac{J N}{2} \left( \hat{\phi}_1^2 + \hat{\phi}_2^2 \right). 
%\label{eq:Hharmonic1}
%\end{eqnarray}
In the lowest-order approximation $cos \hat{\phi}_{A,B} \simeq 1 - \hat{\phi}_{A,B}^2/2$, the system is described by two {\em coupled} harmonic oscillators. This suggests that the system under study can indeed satisfy the entanglement or two-mode squeezing criteria, if the relevant collective variables in our system are mapped onto those of two field modes mixed by a symmetric beam splitter
\begin{eqnarray}
\hat{n}_\pm = \frac{1}{\sqrt{2}} \left( \hat{n}_A \pm \hat{n}_B \right) \leftrightarrow \hat{x}_{\pm}, \nonumber \\
\hat{\phi}_\pm = \frac{1}{\sqrt{2}} \left( \hat{\phi}_A \pm \hat{\phi}_B \right) \leftrightarrow \hat{p}_{\pm}.
\label{eq:nphi_pm}
\end{eqnarray}

Using the collective variables defined in (\ref{eq:nphi_pm}), we can rewrite Eq. (\ref{Hnphi2}) in the harmonic approximation, assuming $(E_c)_{AA} \simeq (E_c)_{BB} = E_c$, as:
\begin{eqnarray}
\hat{H} &=& \left( E_c + (E_c)_{AB} + \frac{2J}{N} \right) \hat{n}_+^2 + \frac{JN}{2} \hat{\phi}_+^2 \nonumber \\
&+& \left( E_c - (E_c)_{AB} + \frac{2J}{N} \right) \hat{n}_-^2 + \frac{JN}{2} \hat{\phi}_-^2.
\label{eq:Hharmonic2}
\end{eqnarray}
Hence, the transformed Hamiltonian describes two {\em uncoupled} harmonic modes in the collective basis. The ``+''-mode corresponds to Josephson oscillations of the total atomic population (regardless of the internal state) between the two wells, such that the inter-species ratio in each well is constant (in-phase oscillations of the $A,B$ species). The ``-''-mode corresponds to oscillations of the inter-species ratio between the two wells, such that the total population imbalance does not change (out-of-phase oscillations of the $1,2$ species).
These two modes have different fundamental frequencies, $\omega_{\pm}$ (see Supplement).

We may then wonder: do the EPR correlation criteria hold in the uncoupled ($\pm$ modes) basis?
%It can be checked (supplement) that 
Indeed, they do: for $(E_c)_{AB} > 0$ Eqs. (\ref{eq:2mode_crit}),(\ref{eq:nphi_pm}),(\ref{eq:Hharmonic2}) allow the $\pm$ modes to satisfy the EPR criteria, yielding $s=[(2J/N + E_c + (E_c)_{AB})]/[(2J/N + E_c - (E_c)_{AB})]$. We then obtain $s >> 1$ for $E_c \simeq (E_c)_{AB} >> 2J/N$ and the ground states of both modes, approaching the {\em ideal} EPR limit $s \rightarrow \infty$ of full CV entanglement. Thus, the fact that there is coupling between the {\em original} ($A$ and $B$) modes {\em suffices} to create {\em EPR} correlations between the collective $\pm$ modes, although there is no coupling in the latter basis.

Beyond the lowest-order approximation that has led to (\ref{eq:Hharmonic2}), there is parametric coupling of the collective modes that may induce nontrivial dynamics of CV wavepackets: the slow, $-$, mode can be ``frozen'' at a low-temperature state, while the fast, $+$, mode may be kept at its ground state, conforming to the Born-Oppenheimer coupling regime (see Supplement). The occupations of thermally excited $+$ mode states must be low compared to its ground state, in order to satisfy the EPR criteria (see Supplement).

For {\em exact} calculation of the dynamics we must resort to the angular momentum operators that describe the full system (Supplement). The entanglement criterion is then\cite{polzik_nature}
\begin{equation}
\frac{1}{| \langle \hat{L}_x \rangle |^2} \left( \langle \Delta \hat{L}_{y \pm} ^2 \rangle \langle \Delta \hat{L}_{z \mp} ^2 \rangle \right) \equiv \frac{1}{4s}  <  \frac{1}{4}.
\label{L_crit}
\end{equation}
This entanglement criterion differs from those used for the number-phase operators only for significant nonlinear phase diffusion, which reduces $| \langle \hat{L}_x \rangle |$ compared to $1$ and thus diminishes the ideal limit of $s$. Since such phase diffusion occurs due to the interatomic (nonlinear) interaction, which is also responsible for the entanglement, one needs to find the optimal charging energies in (\ref{Hnphi2}) and state-preparation that would yield the largest EPR correlations (see Methods).

The optimal {\em sudden} sequence for state preparation consists of (Fig. 1a): (a) filling the original trap by a BEC in internal state $| A \rangle$; (b) sudden ramping up of the inter-well potential barrier, thus creating a two-well symmetric superposition; (c) transforming state $| A \rangle$ into a symmetric superposition of $| A \rangle$ and $| B \rangle$ by a fast $\pi/2$ pulse. This sequence yields an initial {\em coherent state in the two original modes $A$,$B$}, whose EPR entanglement then builds up with time according to their coupled-pendula dynamics (Eq. (\ref{Hnphi2})).
By contrast, {\em slower} ramping up of the barrier causes them to be exposed to both nonlinear phase diffusion and environment-induced dephasing (see below) much longer, thus spoiling the entanglement criterion (\ref{L_crit}) (Fig. \ref{fig:squeezing}(b),(c)). We find an optimal value for the charging energy which results in the largest amount of EPR correlations, closest to the ideal inseparability (obtainable in the absence of nonlinear phase diffusion and for the ground-state of the coupled two-mode system).

We note that it is not advantageous in this scheme to create a single-mode squeezed state in each well as an initial condition. Intuitively, this is due to the fact that such squeezing does not translate into correlations between the wells, and thus does not induce reduced variances of the two-mode coordinates. In more detail, an initial coherent state provides minimal non-correlated variances in the combined variables $\langle \Delta \hat{L}_{y \pm} \rangle \langle \Delta \hat{L}_{z \pm} \rangle / \left| \langle \hat{L}_x \rangle \right|^2 = 1/4$. In comparison, an initial single-mode squeezed state in each well, which is squeezed along the same quadrature, would not improve on this variance product. Finally, initial single-mode squeezed states of different quadratures would result in a larger variance product, thus limiting the two-mode squeezing reachable through the dynamics described above (see Suppl).

\begin{figure}
\subfigure[]{
\includegraphics[width=0.48\linewidth,height=0.35\linewidth]{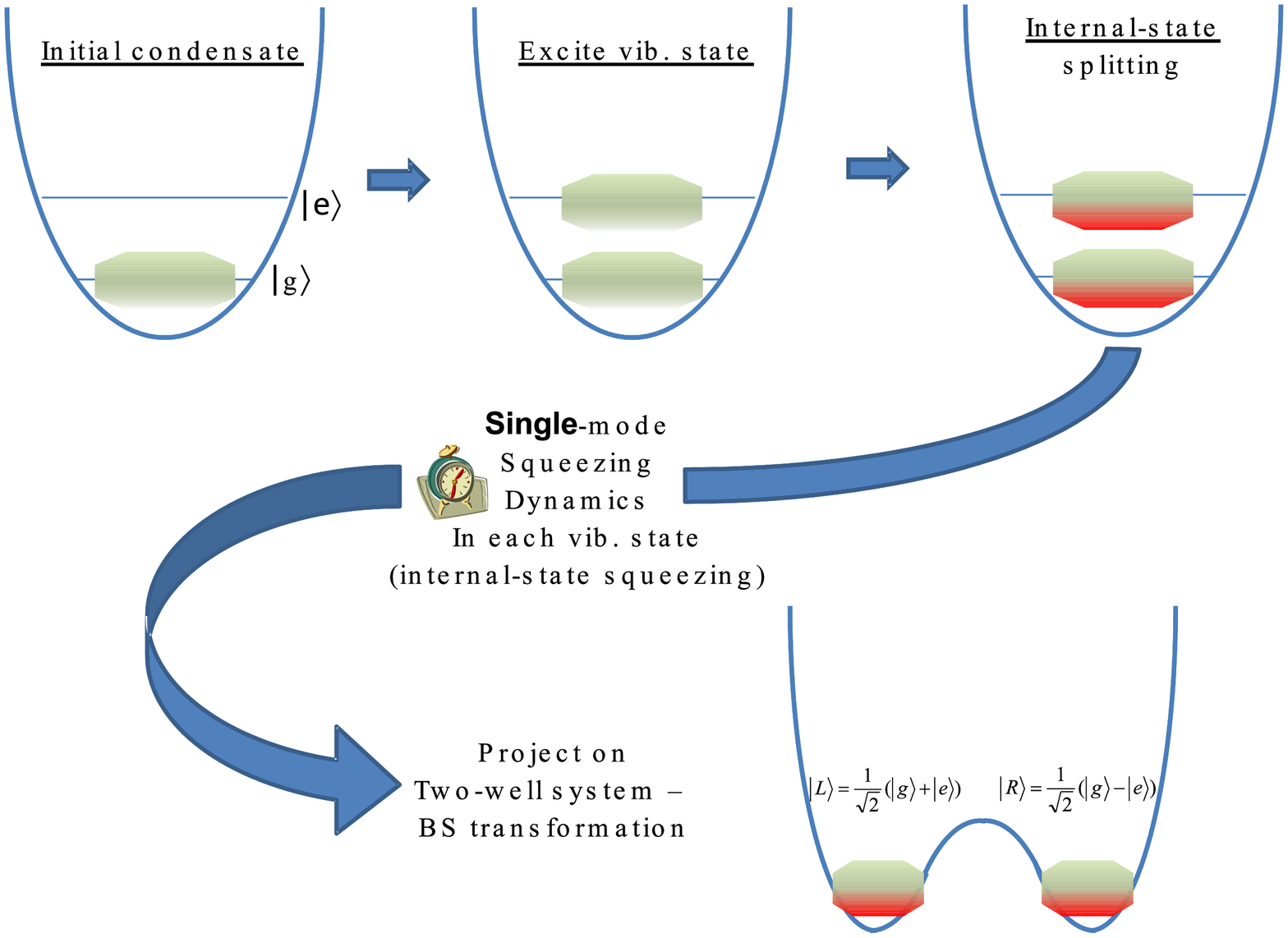}}
\subfigure[]{
\includegraphics[width=0.48\linewidth,height=0.35\linewidth]{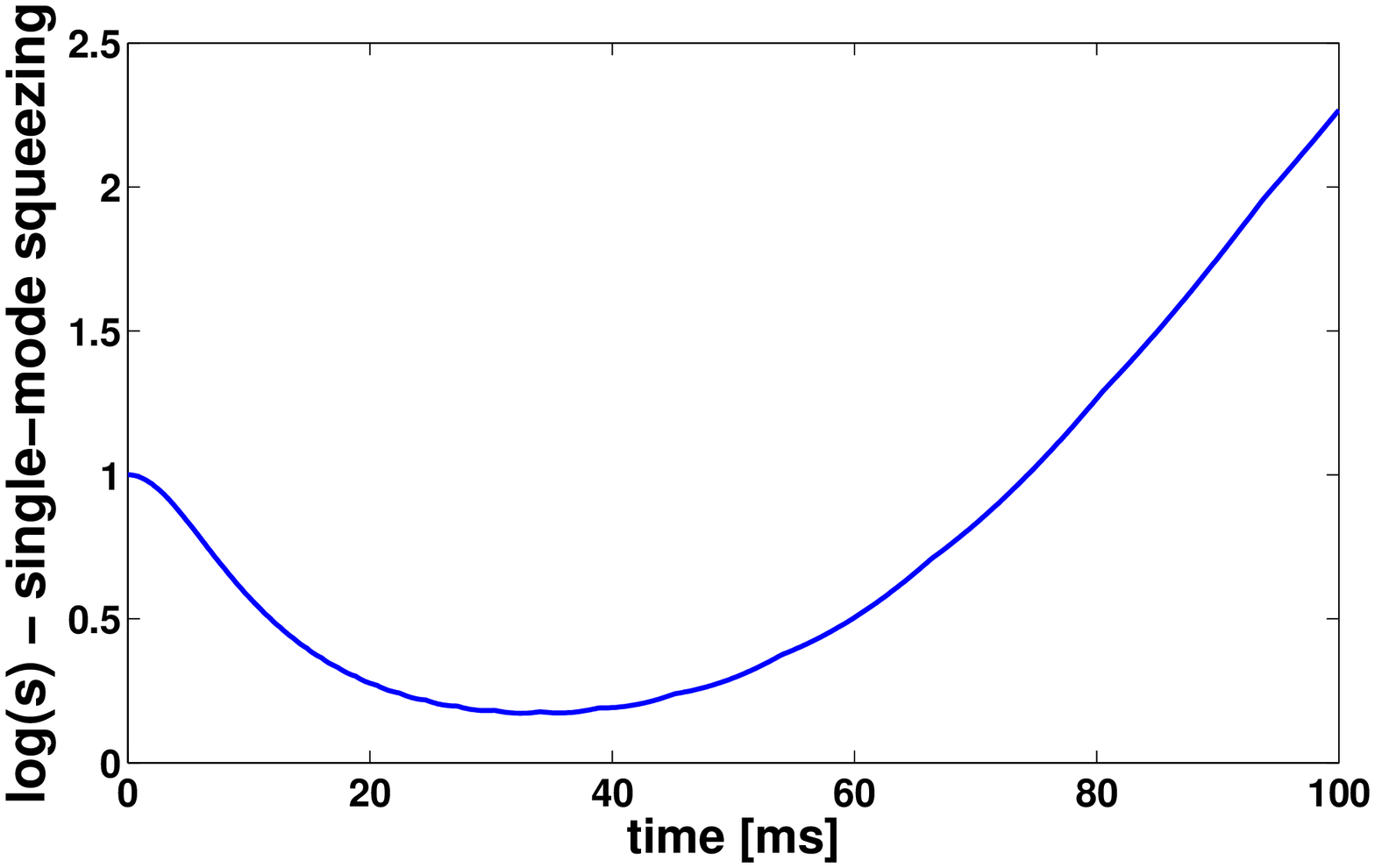}}
\subfigure[]{
\includegraphics[width=0.48\linewidth,height=0.35\linewidth]{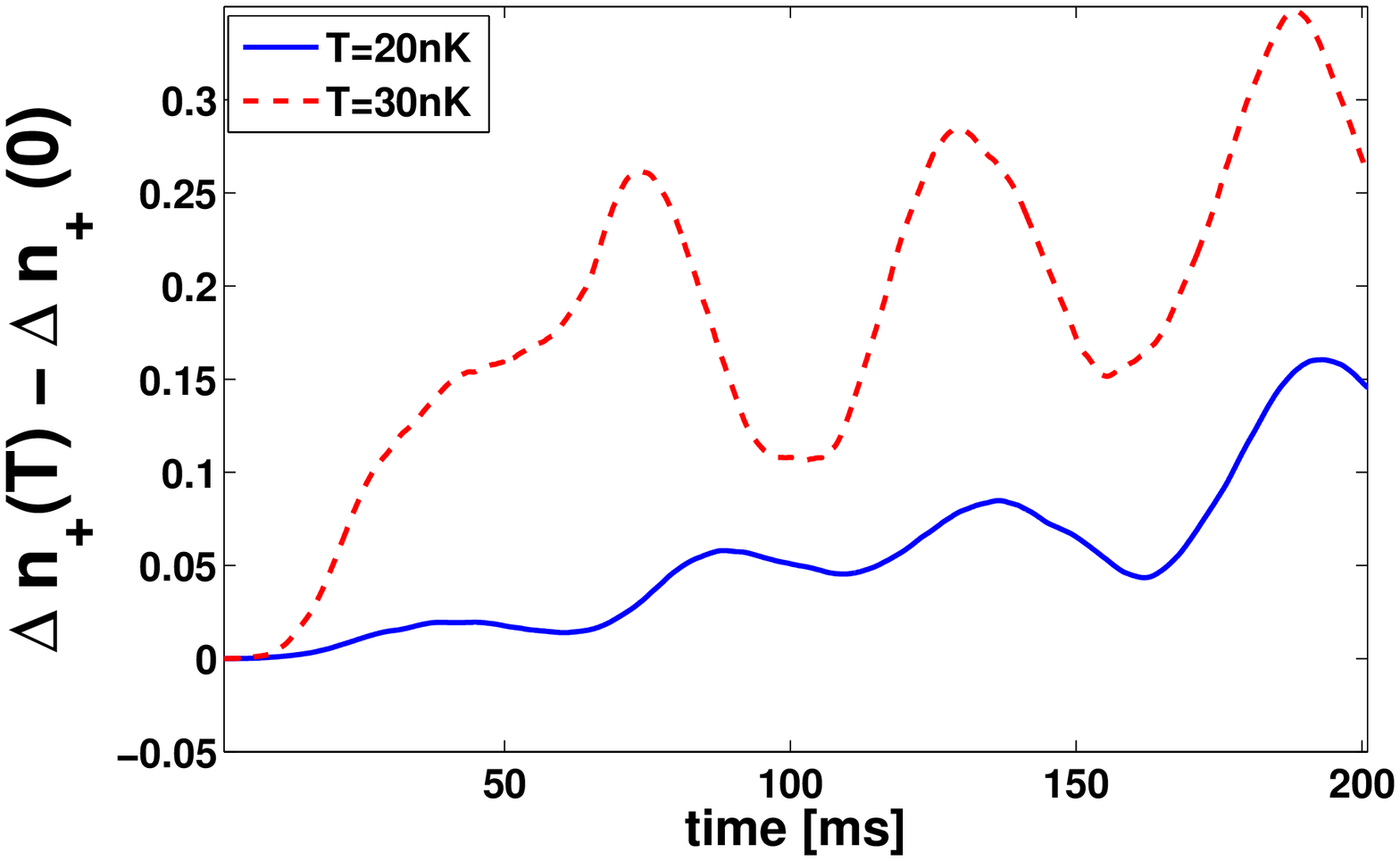}}
\subfigure[]{
\includegraphics[width=0.48\linewidth,height=0.35\linewidth]{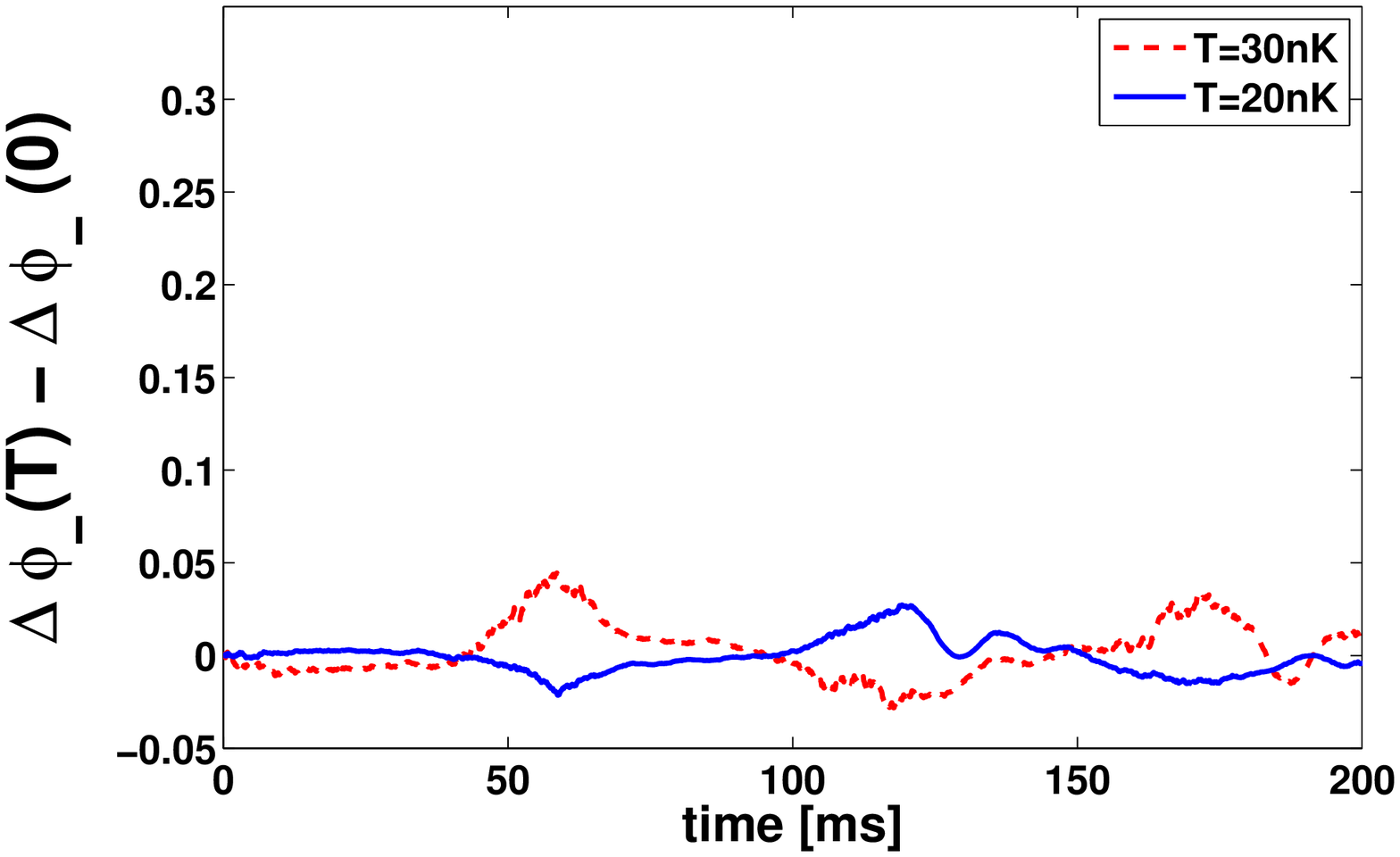}}
\caption{(a) Schematic sequence for the creation of ``non-local'' two-mode entanglement in analogy with the BS approach. (b) Single-mode squeezing dynamics as a function of time, for $N=100$.
Decoherence effects: the variance of $n_+$ (c) and of $\phi_-$ (d) as a function of time, in the presence of proper dephasing. We subtract the variance of the Hermitian dynamics (without dephasing) in order to single-out the effect of dephasing on the dynamics of the variances. The coherence time is here estimated to be $\sim 100$ ms at $20$ nK.}
 \label{fig:BS_new}
\end{figure}

\paragraph*{Scheme for local-mode correlations and ``steering'' in bosonic JJs --}
We now present an approach based on correlations of two squeezed {\em local} (left- and right- well) modes (Fig. \ref{fig:BS_new}(a). The system is initialized in the left well (L) of a double-well potential, in a single internal state ($1$). Then, the barrier is suddenly dropped in order to create a coherent superposition of the vibrational ground-state $| g \rangle$ and first excited state $| e \rangle$ of the new potential.
Next, a $\pi/2$ pulse creates a coherent superposition of the internal states $A,B$ of the atoms.

We assume that there is no exchange term, since to lowest order the cross-coupling terms cancel, and therefore the number of particles in each external state is conserved. This conservation allows us to rewrite the Hamiltonian in terms of the internal-state number difference operator in each vibrational state, $\hat{n}_{g} = (\hat{n}_g)_A - (\hat{n}_g)_B$ and $\hat{n}_{e} = (\hat{n}_e)_A - (\hat{n}_e)_B$. The Hamiltonian then becomes (see Supplement):
\begin{equation}
\hat{H} = \left( (E_c)_{AA} + (E_c)_{BB} - 2 (E_c)_{AB} \right) \left( \hat{n}_{g}^2 + \hat{n}_{e}^2 \right).
\label{eq:1mode_sq}
\end{equation}
Thus, the system evolves {\em separately} in the two vibrational modes, each undergoing dynamical single-mode squeezing in the internal-state basis\cite{ueda,sorensen}. Such internal-state squeezing in each mode was demonstrated recently \cite{oberthaler_squeezing}.

Following an evolution during which each vibrational mode separately experiences internal-state squeezing, we raise the barrier quickly to create two separate symmetric wells, denoted L (left) and R (right). This sudden projection creates a BS-like transformation:
\begin{eqnarray}
|L \rangle &=& \frac{1}{\sqrt{2}} \left( |g \rangle + |e \rangle \right ), \nonumber \\
|R \rangle &=& \frac{1}{\sqrt{2}} \left( |g \rangle - |e \rangle \right ).
\end{eqnarray}
Therefore, we now have two-mode squeezing, or EPR-like entanglement, between the left and right wells. The mode in each well is defined by the number and phase differences of the internal states. Local measurements may be done in the internal-state basis in each well {\em separately}, exhibiting {\em non-classical} correlations between the $|L \rangle$ and $|R \rangle$ spatially separated modes, in the spirit of ``steering''.

The scheme presented above is analogous to the quantum optics approach\cite{braunstein_rev}, in which two independent single-mode squeezed states are injected into the input ports of a beam splitter (BS), thereby creating entangled modes at the output ports of the BS. 
However, the {\em intrinsic} nonlinearity of each BEC mode causes their unwarranted mixing even {\em before} the BS-like transformation, causing fidelity loss (see Methods).

In this sequence we wait for the {\em maximal} single-mode squeezing to develop separately, before raising the barrier to project the $|g \rangle$ and $|e \rangle$ states onto the $|L \rangle$ and $|R \rangle$ basis. Therefore, we can immediately use the maximal squeezing factor $s$ calculated for each single-mode\cite{braunstein_rev}, to extract the two-mode squeezing parameter.
Then the collective two-mode squeezing is given by
\begin{eqnarray}
\langle \Delta \hat{n}_+^2 \rangle &=& \langle \Delta \left( \hat{n}_L + \hat{n}_R \right)^2 \rangle = \frac{ \left \langle \Delta \left( n_+^{(0)} \right)^2 \right \rangle}{s}, \nonumber \\
\langle \Delta \hat{\phi}_-^2 \rangle &=& \langle \Delta \left( \hat{\phi}_L - \hat{\phi}_R \right)^2 \rangle = \frac{\left \langle \Delta \left(\phi_-^{(0)} \right)^2 \right \rangle}{s},
\end{eqnarray}
namely, the two-mode squeezing parameter is equal to that of single-mode squeezing. This squeezing parameter now characterizes the knowledge obtained about variables in one well having measured their counterparts in the other well.

\paragraph*{Decoherence effects --}
We now turn to the effect of environment-induced decoherence on the robustness of EPR entanglement in this system. We assume proper dephasing created by independently fluctuating (stochastic) energy shifts of atoms in each internal state and well $1(2)l(R)$, caused by the thermal atomic or electromagnetic environment. Due to the spectroscopic similarity of the two BEC species, we reduce the number of independent stochastic processes by setting $\epsilon_{1L}/\epsilon_{2L} = \epsilon_{1R}/\epsilon_{2R} = \left(1 - \xi \right)/\left(1 + \xi \right)$, and assuming a {\em "symmetrized environment"}, i.e. $\xi << 1$.
Due to the small value of $\xi$, the variance of $\hat{\phi}_-$ almost does not change (in either the global or local scheme), while the variance of $\hat{n}_+$ increases linearly, and is responsible for the growing loss of entanglement.
Hence, we may manipulate the system as we see fit {\em within} the coherence time. 

%\section*{Experiment:}

\paragraph*{Discussion --}
We have addressed EPR effects in an ultracold-atom analog of two {\em coupled} Josephson junctions (JJs): a {\em two-species} Bose-Einstein condensate (BEC), each species corresponding to a different sublevel of the atomic internal ground state \cite{lobo}, trapped in a tunnel-coupled double-well potential (Fig. 1a). We have shown that such bosonic coupled JJs can induce EPR entanglement of appropriate combinations of collective {\em continuous} (phase and atom-number) variables. This entanglement has been shown to be resilient to environmental noise (decoherence). It exhibits intriguing dynamics under conditions analogous to the {\em molecular} Born-Oppenheimer regime for coupled slow and fast variables \cite{Davydov}. Alternatively, it can dynamically realize beam-splitter mixing of two squeezed modes.

We acknowledge the support of GIF, DIP and EC (MIDAS STREP, FET Open), and the Humboldt Foundation (G.K.).

%\rule{0.5\linewidth}{.2pt}
\bibliography{bibBEC3}

\end{document}